# Improving Time Estimation by Blind Deconvolution: with Applications to TOFD and Backscatter Sizing

Roberto H. HERRERA [1], Zhaorui LIU [1], Natasha RAFFA [1], Paul CHRISTENSEN [1], Adrianus ELVERS [1]

[1] UT Technology a RAE Energy Company; Edmonton, Canada
Phone: +1 780 930 2802, Fax: +1 780 484 5664; e-mail: hherrera@uttechnology.com, rliu@uttechnology.com, natasha@uttechnology.com, paul@uttechnology.com, adrianus@uttechnology.com

**Abstract**
In this paper we present a blind deconvolution scheme based on statistical wavelet estimation. We assume no prior knowledge of the wavelet, and do not select a reflector from the signal. Instead, the wavelet (ultrasound pulse) is statistically estimated from the signal itself by a kurtosis-based metric. This wavelet is then used to deconvolve the RF (radiofrequency) signal through Wiener filtering, and the resultant zero phase trace is subjected to spectral broadening by Autoregressive Spectral Extrapolation (ASE). These steps increase the time resolution of diffraction techniques. Results on synthetic and real cases show the robustness of the proposed method.

## 1. Introduction

In ultrasonic testing, the emitted ultrasound pulse $w(t)$ is reflected or diffracted at the discontinuities or density changes in the medium. The location in time and amplitude of these reflection coefficients builds up the reflectivity function $r(t)$, which is the ultimate goal in the inversion process [1]. The interaction of the pulse with each discontinuity is mathematically represented by the convolutional model [2]:

$$s(t) = w(t) * r(t) + \eta(t), \qquad (1)$$

where $s(t)$ is the measured signal, the operator * denotes the convolution operation and $\eta(t)$ is the additive noise always present.

While we use the ultrasound pulse $w(t)$ to get information from the medium $r(t)$, this very same pulse affects the resolution of our measuring system. This is because the pulse is band-limited and therefore far from the desired Dirac delta. In order to recover the desired spiky impulse response $r(t)$ from the band-limited observation $s(t)$, we must remove the pulse effects. This process is called deconvolution, and it is aimed at increasing the time resolution of (ultrasonic) UT acquisition system.

From equation 1 above, we know that the spectrum in the passband can be obtained by simple deconvolution [3]. Therefore, the reference wavelet plays a key role in the quality of the reconstructed impulse response. This reference wavelet is usually deterministically estimated from the observation itself by selecting a representative segment of the signal, such as the backwall reflector [4].

Another deterministic method is to measure the reference wavelet in a pulse-transmission setup [5]. However, the time-varying nature of the ultrasound pulse makes these estimates only locally correct. TOFD signals pose an extra challenge due to the weak diffracted signals and angular variations of the pulse shape. Honarvar and Yaghootian [5] propose a local deconvolution alternative, which involves segmenting the TOFD signal into quasi-stationary regions and deconvolving with a dictionary of angle-dependent reference wavelets. They show how important the choice of the wavelet is for an accurate deconvolution. However, it is a labour intensive task to setup a custom dictionary of reference wavelets for every situation and different inspection conditions.

Yet, there is a third robust way to estimate the wavelet based only on statistical signal processing methods [6]. Statistical wavelet estimation has been successfully used in seismic deconvolution (e.g., [7], [8], [9]) and in ultrasonic signals (e.g., [6], [10]). When both the pulse *w(t)* and the reflectivity *r(t)* are unknowns, we say that it is a blind-deconvolution process [11]. This is because the reference wavelet has to be estimated beforehand, and statistical methods are robust tools in this case. We assume that the reflectivity is white and thus its autocorrelation function is a delta at zero-lag [12]. This is a reasonable assumption due to the sparse nature of the ultrasound signal where reflectors are randomly distributed and separated in time. From the entire signal only a few points are actual reflectors.

The vast majority of deconvolution methods rely on the whiteness assumption (e.g., [1], [12], [13]). Under this constraint, we can directly estimate the autocorrelation of the wavelet from the autocorrelation of the observed ultrasound trace. However, the autocorrelation is a second order statistics and carries no phase information. We still need a method to reconstruct the inverse filter with phase information to yield a zero-phase deconvolved signal. Van der Baan [7] proposed a statistical approach, based on kurtosis maximization, to estimate the phase of a seismic wavelet from data alone. In this paper we extend this approach to ultrasonic signals.

The wavelet spectrum is computed from the autocorrelation of the trace and its phase is estimated through a search of constant-phase rotations while computing the kurtosis [7]. The rationale behind this approach is that convolving a filter (ultrasound wavelet) with a white time series (reflectivity) renders the outcome more Gaussian [7]. Thus, the desired white reflectivity series can be recovered by deconvolving the output *y(t)* with an inverse filter that ensures the deconvolution output is maximally non-Gaussian [7], [14].

In this paper, we apply the statistical wavelet estimation method to deconvolve diffracted signals from TOFD and back-diffraction experiments. We extend the method by computing a wavelet for each trace and deconvolving by Wiener filtering, followed by autoregressive spectral extrapolation. The wavelet estimation step is done in each ultrasound trace instead of using one unique wavelet for an entire TOFD image. The same approach is used in sectorial scans (S-Scans) to deconvolve the back-diffracted signals.

## 2. Statistical wavelet estimation

It is common practice in the deconvolution of ultrasonic signals to use one of the echoes of the received signal, such as the backwall echo, as the reference wavelet [4], [5]. However, this practice has an inherent disadvantage: reflections with shapes that differ from the preselected reflector will be misrepresented or totally ignored by the deconvolution process.

Statistical tools have shown to be more robust for wavelet extraction in geophysics [7], [8]. Here we apply the wavelet estimation based on kurtosis maximization to ultrasonic signals.

### 2.1  Wavelet estimation by kurtosis maximization

Maximization of the kurtosis by constant phase rotation of the observed ultrasound trace can reveal the phase of an ultrasonic wavelet. The rationale behind this method is that convolving any white reflectivity with an arbitrary wavelet renders the outcome less white, but also more Gaussian [14]. Since kurtosis measures the deviation from Gaussianity, we can recover the original reflectivity by maximizing the kurtosis [7].

The rotated ultrasound trace $s_{rot}(t)$ can be computed from the observed trace *s(t)* by

$$s_{rot}(t) = s(t)\cos\Phi + H[s(t)]\sin\Phi, \quad (2)$$

with $\Phi$ indicating the phase rotation angle and H[.] indicating the Hilbert transformer operator. At each angle rotation we measure the deviation from Gaussianity using the kurtosis, calculated as:

$$\kappa[s] = \frac{E[s^4]}{(E[s^2])^2} - 3, \quad (3)$$

where E[.] denotes the expected value. When the kurtosis value reaches its maximum, we get the desired wavelet phase; i.e., it is equal to $-\Phi_{kurt}$, the rotation angle at the maximum kurtosis value [7]. Finally, the wavelet is estimated by inverse Fourier transform of the following equation:

$$W_{est}(f) = |S_{cc}(f)|\exp\{i\,\Phi_{kurt}sgn(f)\}, \quad (4)$$

where $|S_{cc}(f)|$ is the magnitude spectrum of the observed signal computed from the autocorrelation function of *s(t)*, and $sgn(f)$ is the sign correction for positive and negative frequencies.

## 3. Wiener filtering

Once the reference wavelet *w(t)* is estimated we can compute the Wiener filter *g(t)*, which is the optimum inverse filter in the least square sense [15]. In the frequency domain, it is given by:

$$G(f) = \frac{W^*(f)}{|W(f)^2| + \sigma_n^2}, \quad (5)$$

with $\sigma_n^2$ the noise variance, and superscript * stands for complex conjugate [7].

The regularization parameter in the denominator is often called noise desensitizing factor and it could be substituted by $\sigma_n^2 = \max(|W(f)|^2)/100$ [4]. Then, the estimated reflectivity in the frequency domain is the spectrum of the observed signal *S(f)* filtered by *G(f)*:

$$R_{est}(f) = \frac{S(f)W^*(f)}{|W(f)^2| + 0.01\max|W(f)^2|}. \quad (6)$$

The inverse Fourier transform of $R_{est}(f)$ gives the zero phase estimated reflectivity.

## 4. Autoregressive spectral extrapolation

The output of the Wiener filter is a zero-phase signal with its maximum values located at the correct reflector positions. The band-limited nature of the wavelet used for deconvolution makes the pulse

wide— which in turn, makes it hard to identify the exact reflector location. This could be solved by spectral broadening using autoregressive techniques that were originally developed for seismic signal processing [16], [17] and later extended to ultrasonic signals [3], [18].

Autoregressive spectral extrapolation (ASE) has attracted the interest of the non-destructive testing community as a resolution enhancement solution for defect sizing ([4], [19], [20], [21]). Honarvar et al. [4] settled the basis for an extensive use of this method in further studies. Herrera et al. [6] integrated the ASE method into a blind deconvolution framework using a Wiener filtering step in the Fourier and wavelet domain. More recently, Sinclair et. al. [19] and Shakibi et.al. [20] used ASE to improve the time resolution in diffraction-based methods.

The rationale behind this method is that the part of the deconvolved spectrum with high signal-to-noise ratio (SNR) can be modelled as an autoregressive (AR) process. The remaining part of the spectrum is then extrapolated based on this region [4], [16].

We use the Burg's method to extrapolate the spectrum of the deconvolved signal. Each point in the selected spectral window, which we chose to be the -6 dB bandwidth (Full width at half maximum - FWHM), can be represented as a linear combination of the $p$ previous and posterior points, where $p$ is the order of the Burg's (autoregressive) AR model. In our implementation we use the Akaike Information Criterion [22] to select the optimum $p$ value.

Once the prediction coefficients are computed using Burg's method, the left part of the spectrum (0 to $m$) and right part ($n$ to Nyquist frequency, $N$) are extrapolated using the linear prediction equations:

$$\hat{X}_l(f) = -\sum_{i=1}^{p} a_i^* X_{l+i}(f) \quad l = 0, 2, \ldots, m-1 \quad (7)$$

$$\hat{X}_r(f) = -\sum_{i=1}^{p} a_i X_{r-i}(f), \quad r = n+1, n+2, \ldots, N \quad (8)$$

where $\hat{X}_l(f)$ and $\hat{X}_r(f)$ are the backward and forward linear predictions respectively and $a_i$ and $a_i^*$ are the prediction coefficients and their complex conjugates.

## 5. Synthetic example

In this synthetic example, we integrate into one hypothetical RF signal the possible responses from TOFD and the back diffraction experiments. Figure 1a shows the TOFD experimental setup, and Figure 1b shows the back diffraction technique for a surface-breaking flaw.

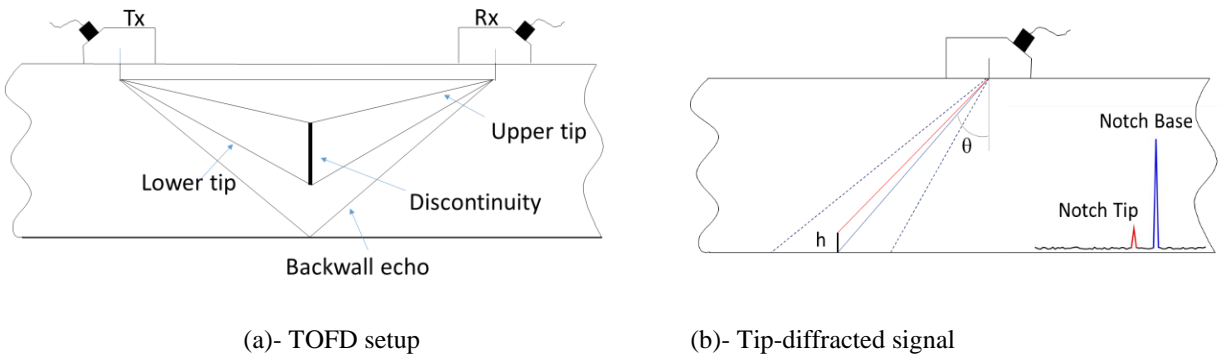

(a)- TOFD setup  (b)- Tip-diffracted signal

Figure 1. Synthetic example, experimental setup. a) The TOFD method and b) the back diffraction technique.

Both methods are crack tip diffraction techniques with TOFD using the forward propagated signals in a pitch-catch setup. The back diffraction method makes use of the backscattered signal in a single transducer (pulse-echo) configuration. Resolution is the capacity to separate close-spaced reflectors generated at the crack tips. To have a good separation in time, we need a broad-band wavelet which is only achieved by deconvolution.

Figure 2 depicts the hypothetical reflectivity function comprising both experiments. The reflectivity is the sum of all of the reflection coefficients, represented by deltas positioned at the reflector location. Our goal is to recover this reflectivity from the observation, keeping the phase and magnitude of these reflectors.

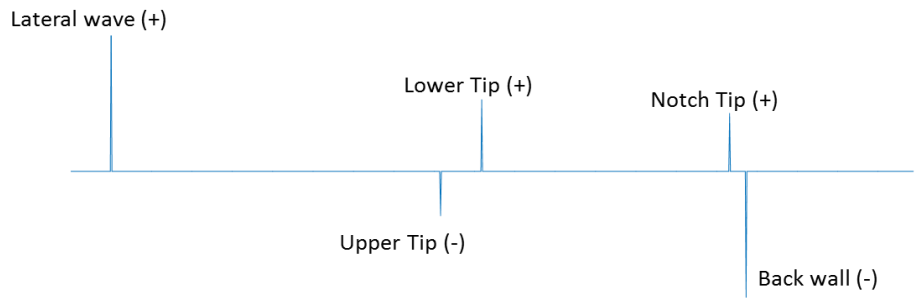

Figure 2. Hypothetical reflectivity. Note the phase signs at each reflector. From left to right, the first wave to arrive is originated at the emitter transducer with positive sign. In the TOFD experiment upper tip (-) and the lower tip (+), note that the separation of these reflectors is proportional to the crack size. The delta (+) corresponding to a surface-breaking flaw is proportional to the notch size. Finally, the strong backwall reflector shows negative polarity.

Following equation 1, and convolving a generated ultrasonic pulse (wavelet $w(t)$) with the reflectivity $r(t)$ in Figure 2, we get the observed RF signal. To simulate real signals, we add some Gaussian noise to have a signal-to-noise ratio (SNR) of 15 dB. Figure 3 shows the entire convolution process along with the estimated wavelet (top plot in red).

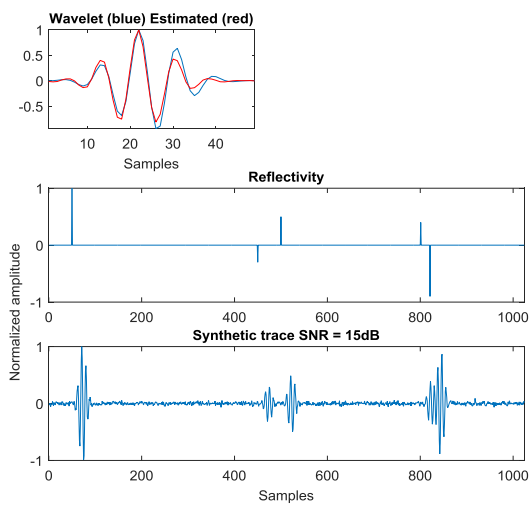

Figure 3. Generating the observed RF signal by convolution. The ultrasound pulse (top) is convolved with the reflectivity (center) to generate the RF signal (bottom). The estimated wavelet from the noisy signal is shown in red in the top plot.

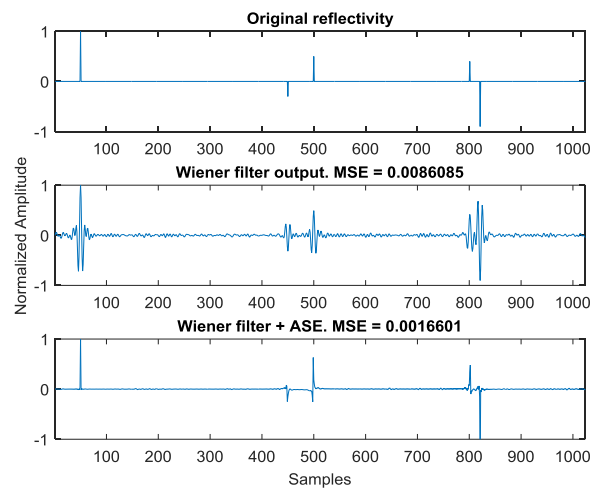

Figure 4. Deconvolution results. The original reflectivity (top) shows the ideal magnitude and phase for each reflector. The Wiener output is comprised of zero phase wavelets placed at the reflectors' locations. The bottom plot shows the results after spectral broadening using ASE.

The output of the Wiener filter is shown in the center plot of Figure 4. Note that the original RF turned into a zero-phase signal with the maxima placed at the reflectors' locations. The importance of converting a signal to zero-phase is that these type of signals have the maximum instantaneous energy and hence the best resolving power. These properties improve detectability, which is the ability to be seen in the presence of noise [23].

Figure 5 shows the improvement in resolution produced by the ASE method. The convolution of a band limited wavelet (ultrasonic pulse) with infinite bandwidth deltas (reflectivity coefficients) produces a band-limited output. Thus, the spectrum of the signal (in blue) is also band-limited. The Wiener filter only flattens the spectral region, while the goal is to whiten the spectrum. Extrapolating the "good" spectral region involves approximating the spectrum to the one that corresponds to the reflectivity. Comparing the spectrum of the deconvolved signal (Wiener + ASE in red) and the actual reflectivity spectrum (black), we can see a good agreement.

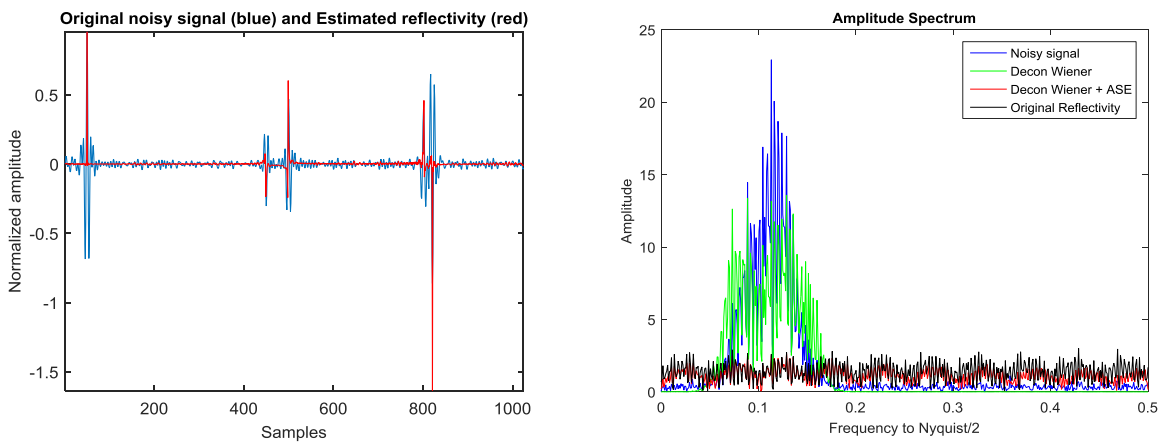

Figure 5. Left plot overlays the deconvolved signal with Wiener + ASE (red) and the Wiener filtered signal (blue). Note that the red signal is able to clearly isolate the close-spaced reflectors. The right plot shows the spectral broadening with the deconvolved spectrum (red) approximating the original reflectivity spectrum (black).

Ignoring the phase information, a better representation can be obtained by finding the envelope of the final deconvolved signal. The envelope of the signal shows the positions of the local maxima at the maximum energy. This helps to improve the location by visually picking the correct maximum. The envelope of the signal is computed as the absolute value of its Hilbert transform (see Figure 6). Note that the initially overlapped reflector from the notch tip is clearly resolved.

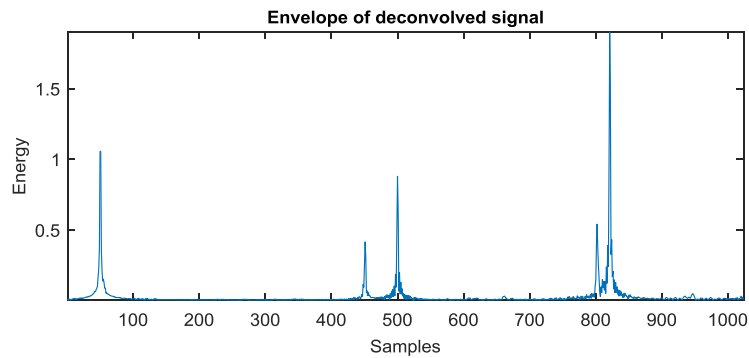

Figure 6. Final deconvolved signal computed as the envelope of the analytical signal using the Hilbert transform. Reflectors are easily identifiable in this signal.

## 6. Real data examples

### 6.1 Deconvolution of TOFD dataset

In this experiment we use a TOFD dataset acquired using the UTScan system at maximum resolution (fs = 100 MS/s). Figure 7a shows the raw data and Figure 7b depicts the deconvolved image using the kurtosis-based method. Events are more localized and sharper in the deconvolved image. We can still see the phase effects on the edges of the diffracted pulses, but we can identify the signals corresponding to the rising edges, which have opposite phases for crack tips and in A-Scan 120.

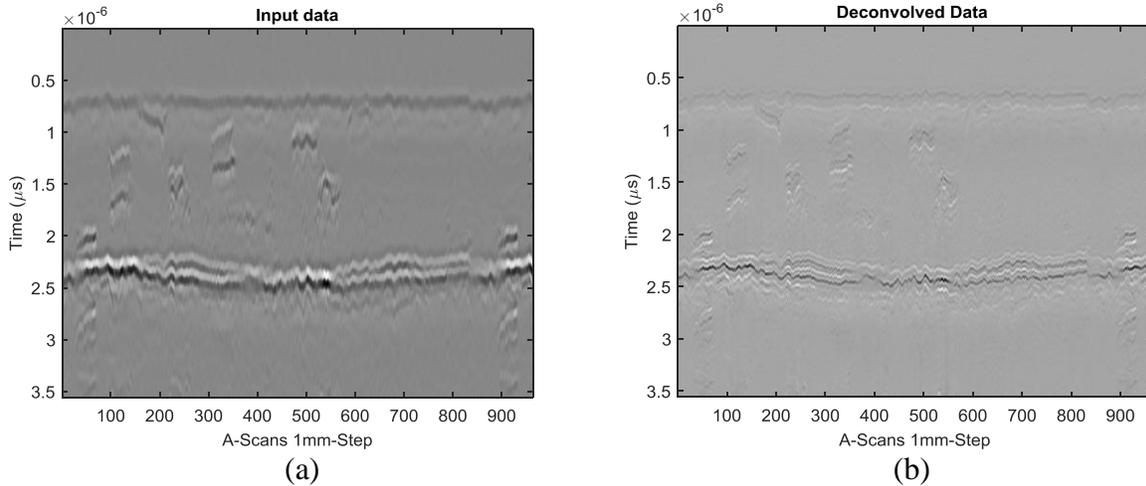

Figure 7. a) TOFD raw data and b) Deconvolved image using KPE. Note the improvement in SNR and the definition of major reflectors.

Figure 8 shows the A-Scan 120 and the deconvolved signal using the estimated wavelet plotted in red. We estimated one wavelet per trace; alternatively, we can estimate a unique wavelet for the entire section or divide the section into quasi-stationary sectors, as proposed by Honarvar and Yaghootian [5]. The deconvolved signal clearly shows the four main components from the A-Scan, namely: the lateral wave (LW), the upper tip (U-Tip) with negative phase and the lower tip (L-Tip) with opposite phase. Finally the backwall reflector (BW) appears around sample 190 with negative phase.

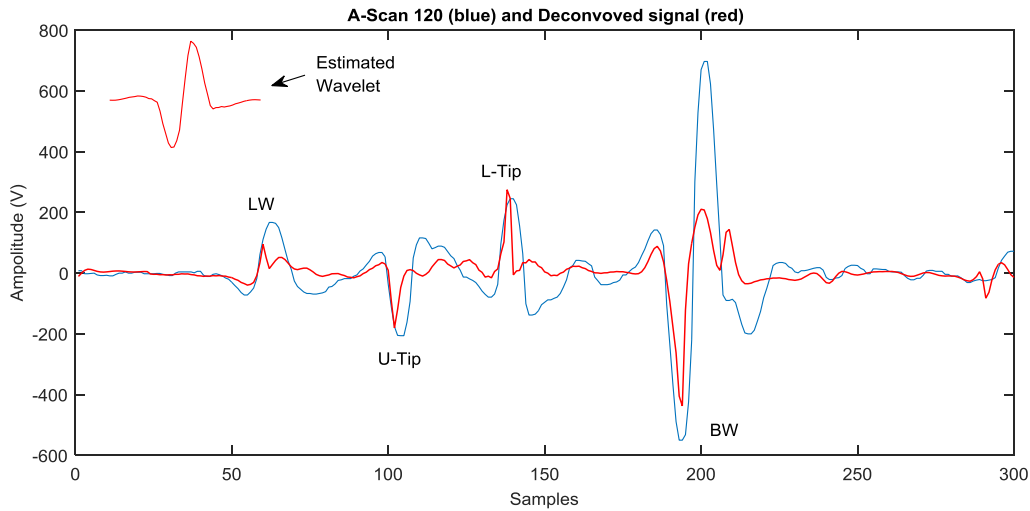

Figure 8. A-Scan 120 in blue and deconvolved signal in red. The estimated wavelet used for the deconvolution as shown at the top left.

## 6.2 Deconvolution of back diffraction dataset

In this experiment, a phased array inspection was carried out to identify the back scattered signal from a surface breaking flaw. Results are presented as a sector scan (S-Scan) in Figure 9 with the top plot showing raw A-scan (shot 15 at an angle of 48.2 degrees), the center plot shows the deconvolved signal and the bottom plot is the envelope of the deconvolved signal.

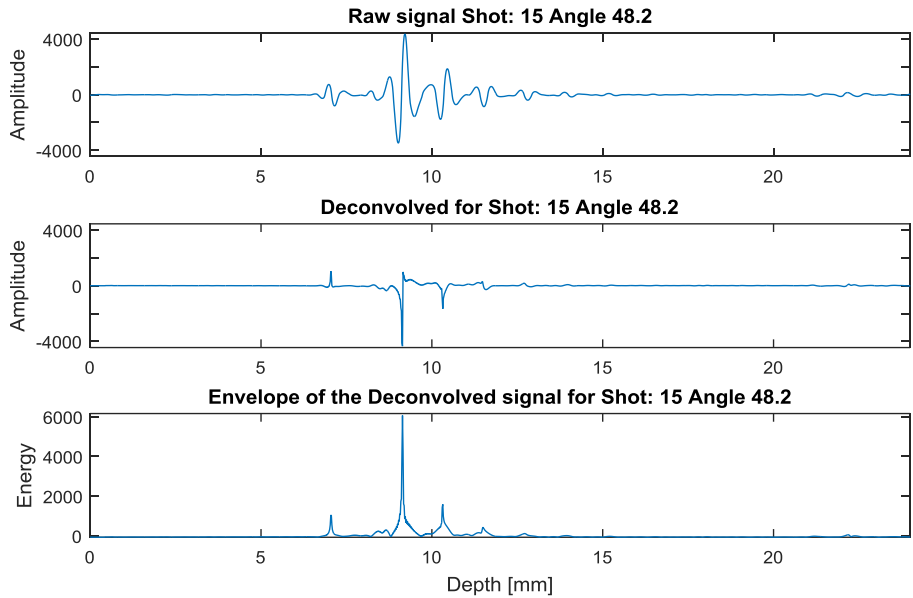

Figure 9. Top plot show the raw RF signal corresponding to shot-15 ($\theta = 48.2$) and the deconvolved signal in the center plot. Note the resolution improvement and the phase identification. The bottom plot is the envelope of the deconvolved signal computed using the Hilbert transform.

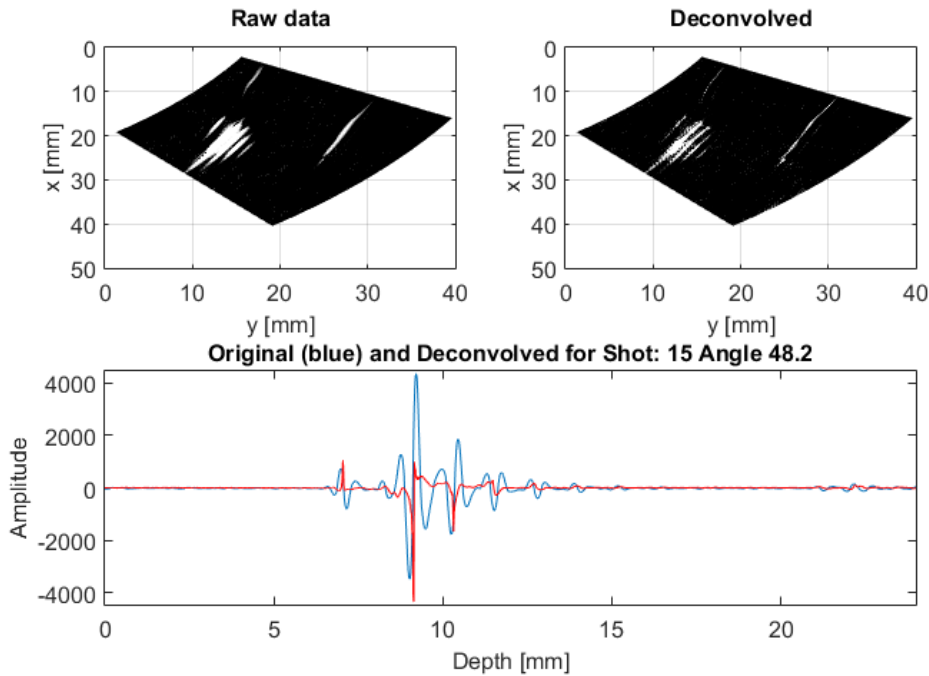

Figure 10. Complete S-Scan for the back diffraction experiment. Top left plot shows the raw data, and the upper right plot shows the deconvolved output. The lower plot shows the raw signal (blue) for Shot 15 at angle 48.2 and the deconvolved signal (red) with the develop method. The crack tip and the back-wall echo are better delineated. Note the time correction in the RF signal and the sharp peaks at each reflector.

The improvement in resolution can be observed in Figure 10. Note that the deconvolved S-Scan clearly delineates the crack tip diffracted signal and separates the back wall reflector from posterior ringing.

## 7. Conclusions and Discussion

Wavelet estimation using the kurtosis maximization method gives good results in isotropic media, where the wavelet changes are mainly due to amplitude attenuation with almost constant frequency. If you expect to have frequency variations, then estimating the wavelet via short-time homomorphic wavelet estimation method [8] should be more appropriate.

Deconvolution can help resolve close-spaced reflectors, and the use of statistical wavelet estimation makes the process more robust to pulse variations due to propagation effects. The deconvolution step can be integrated into the TOFD and S-Scan methods, and it only works over the A-Scan signals. Its implementation could be easily parallelized.

Estimating one wavelet per trace is computationally costly and sometimes not viable for real time processing. In those cases, the proposed method can be adapted to estimate a single unique wavelet for the entire dataset or to estimate localized wavelets for quasi-stationary sections. For example we can divide a TOFD dataset into three zones: lateral wavelet, center and backwall. We can then perform the blind deconvolution method and join the deconvolved sections.

Further work is needed to use the phase information provided by the deconvolution method in both techniques (TOFD and back diffraction).

**Acknowledgements**
The authors would like to thank UT Technology for permission to use the data shown in this paper. We are also indebted to Shealene Page for her review of this manuscript. Henry Herrera thanks Mirko van der Baan for the discussions about statistical wavelet estimation and blind deconvolution.